\documentclass[useAMS]{mn2e} %%% i001
\usepackage{graphicx}

\newcommand{\Tef}{T$_{\rm eff}$~}
\newcommand{\Vm}{$V_{\rm macro}$~}
\newcommand{\Vt}{$V_{\rm t}$~}
\newcommand{\Vexp}{$V_{\rm exp}$~}
\newcommand{\Vmacro}{$V_{\rm macro}$~}
\newcommand{\Vr}{$v*sin\it{i}$~}
\newcommand{\Vg}{$V_{\rm g}$~}
\newcommand{\Vrad}{$V_{\rm r}$~}

\title[
The absorption spectrum of V838 Mon
]{
The absorption spectrum of V838 Mon in 2002 February - March. I.
Atmospheric parameters and iron abundance.\thanks{Based in part on
observations collected with the 1.83m telescope of the Astronomical
Observatory in Asiago, Italy}
}
\author [Bogdan M. Kaminsky, Yakiv V. Pavlenko]{Bogdan M. Kaminsky$^{1}$\thanks{E-mail:
bogdan@mao.kiev.ua}, Yakiv V. Pavlenko$^{1}$\thanks{E-mail:
yp@mao.kiev.ua} \\
$^1$Main Astronomical Observatory of Ukrainian Academy of Sciences,
Golosiiv woods, 03680 Kyiv-127, Ukraine}

\begin{document}

\date{Received ; accepted }

\pagerange{\pageref{firstpage}--\pageref{lastpage}} \pubyear{2004}

\maketitle

\label{firstpage}

\begin{abstract}
We present a determination of the effective temperatures, iron
abundances, and microturbulent velocities for the
pseudophotosphere of V838 Mon on 2002 February 25, and March 2 and
26. Physical parameters of the line forming region were obtained
in the framework of a self-consistent approach, using fits of
synthetic spectra to observed spectra in the wavelength range
5500-6700\AA. We obtained \Tef = 5330 $\pm$ 300 K,  5540 $\pm$ 270
K and 4960 $\pm$ 190 K,  for February 25, March 2, and March 26,
respectively. The iron abundance $\log N(\mbox{Fe})$=$-$4.7 does
not appear to change in the atmosphere of V838 Mon from February
25 to March 26, 2002.

\end{abstract}

\begin{keywords}
stars:atmospheres -- stars:abundances -- stars:individual:V838 Mon
\end{keywords}

\section{Introduction}

The peculiar variable star V838 Mon was discovered during an
outburst in the beginning of 2002 January (Brown 2002). Two
further outbursts were then observed in 2002 February (Munari et
al. 2002a; Kimeswenger et al. 2002; Crause et al.2003) and in
general the optical brightness in V-band  of the  star
increased by 9 mag. Since 2002 March, a gradual fall in
V-magnitude began which, by 2003 January, was reduced by 8 mag.
The suspected progenitor of V838 Mon was identified by
Munari et al. (2002a) as a 15 mag F-star on the main sequence.
Possibly V838 Mon might have a B3V companion (Desidera \&
Munari 2002), but it could be a background star. The discovery of
a light echo (Henden et al. 2002) allowed an estimate of the
distance to V838 Mon and, according to recent works based on HST
data (Bond et al. 2003; Tylenda  2004) its distance is 5-6
kpc. If these estimations are correct, at the time of maximum
brightness V838 Mon was the most luminous star in our Galaxy.

Details of the spectral evolution of the star are described in
Kolev et al. 2002; Wisniewski et al. 2003; Osiwala et al. 2002).
During  outbursts (except for the last) the spectrum
displayed numerous emission lines with P Cyg profiles, formed in
the expanding shell and around an F- or A-star (Kolev et al.
2002). On the other hand, absorption spectra appropriate to a red
giant or supergiant were observed in quiescent periods. Strong
lines of hydrogen, D lines of sodium, triplets of calcium and
other elements show P Cyg profiles. They have similar profiles and
velocities  varying from $-$500 km s$^{-1}$ in late January
to $-$280 km s$^{-1}$ in late March (Munari et al. 2002a). Since
the middle of 2002 March, the emissions are considerably
weakened and the spectrum of V838 Mon evolved to later spectral
classes. In middle of  2002 April, there were present some
lines of TiO; in May the spectrum evolved to the``very cold''
M-giant (Banerjee \& Ashok 2002). In October Evans et al.
(2003) characterized it as a L-supergiant.

Recently Kipper et al.(2004) found for  iron group elements
[m/H] = $-$0.4, while abundances of lithium and some s-process
elements are clearly enhanced. This results was obtain using the
static LTE model.

These results are very dependent on the model atmosphere and
spectrum synthesis assumptions.

The nature of the outbursts remains a mystery. Possible explanations
include various thermonuclear processes (very slow nova, flare
post-AGB), and the collision of two stars (Soker \& Tylenda 2003).
Munari et al. (2002a) suggested that V838 Mon is a new type of a
variable star, because comparison with the closely analogous V4334
Sgr and M31 RV has shown  significant enough differences in the
observed parameters.

In this paper we discuss the results of the determination of iron
abundance and atmospheric parameters of V838 Mon. These we
obtained  from an analysis of absorption spectra of V838 Mon on
2002 February 25 and March 2 and 26. The complexity and uniqueness
of the observed characteristics of V838 Mon practically excluded a
definition of the parameters of the atmosphere using conventional
methods, based on calibration on photometric indices, ionization
balance, profiles of hydrogen lines. Indeed, the presence around
the star of a dust shell, and the uncertain determination of
interstellar reddening (from $\it {E_{\rm B-V}}$~=$-$0.25 to $\it
{ E_{\rm B-V}}$~=$-$0.8 Munari et al. 2002a), affects the $\it
{U-B}$ and $\it {B-V}$ colours. Emission in the hydrogen lines
provides severe problems for their application in the estimation
of effective temperature. Moreover, both the macroturbulent
motions and expansion of the pseudophotosphere merges the numerous
lines in wide blends. As a result, a single unblended line in the
spectrum of V838 Mon cannot be found at all, and  any analysis
based on measurements of equivalent widths is completely excluded.

The observational data used in this paper are described in section
2. Section 3 explains some background to our work and some details
of the procedure used. We attempt to determine \Tef, the
microturbulent velocity \Vt and the iron abundance $\log
N(\mbox{Fe})$ in the atmosphere of V838 Mon in the framework of
the self-consistent approach in section 4. Some results are
discussed in section 5.

\section{Observations}

Spectra of V838 Mon were obtained on 2002 February 25 and March 26
with the Echelle+CCD spectrograph on the 1.82m telescope operated
by Osservatorio Astronomico di Padova on Mount Ekar (Asiago), and
freely available to the community from
http://ulisse.pd.astro.it/V838Mon/. A 2 arcsec slit was used with
fixed E-W orientation, producing a PSF with a FWHM of 1.75 pixels,
corresponding to a resolving power close to 20000. The detector
was a UV coated Thompson CCD $1024\times1024$ pixel, 19 micron
square size, covering in one exposure the wavelength range 4500 to
9480\AA\ (echelle orders \#49 to \#24). The short wavelength limit
is set by a 2~mm OG~455 long-pass filter, inserted in the optical
train to cut the second order from the cross-disperser. The
wavelength range is covered without gaps between adjacent echelle
orders up to 7300\AA. The spectra have been extracted and
calibrated using IRAF software running under Linux operating
system. The spectra are sky-subtracted and flat-fielded. The
wavelength solution was derived simultaneously for all 26
echelle orders, with an average r.m.s of 0.18~km~s$^{-1}$. The
8480-8750~\AA\ wavelength range of these Asiago spectra has been
described in Munari et al. (2002a,b).

Another set of  spectra (R $\sim$ 32000) for March 2 was
obtained with the echelle fibre-fed spectrograph on the
1.9-m SAAO telescope kindly provided for us by Dr. Lisa Crause (see
Crause et al. 2003 for details).

\section{Procedure}

To carry out our analysis of V838 Mon we used the spectral synthesis
techniques. Our synthetic spectra were computed in the  framework of
the classical approach: LTE, plane-parallel media, no sinks and
sources of energy inside the atmosphere, and transfer of energy provided
by the radiation field and by convection.

Strictly speaking, none of these assumptions is 100\% valid
in atmosphere of V838 Mon. Clearly we have non-static
atmosphere which may well have shock waves moving trough it. Still
we assumed that in any moment the  structure of model
atmosphere of V838 Mon is similar to model atmospheres of
supergiants. Indeed, temporal changes of the absorption spectra on
the days were rather marginal. Most probably, for this object, we
see only a pseudophotosphere, which is the outermost part of an
expanding envelope. Therefore, our first goal was to determine
whether it is possible to fit our synthetic spectra to the
observed V838 Mon spectra.

At the time of the observations the spectral class of V838 Mon
was determined as K-type (Kolev et al. 2002). Absorption lines in
spectrum of V838 Mon form comparatively broad blends.

\begin{figure}
\begin{center}
\includegraphics [width=88mm, angle=00]{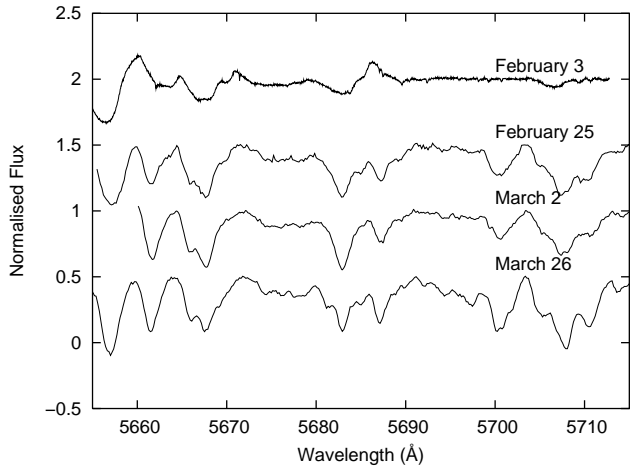}
\end{center}
\caption[]{\label{__sao} Spectra of V838 Mon observed on February
3, February 25, March 2 and March 26 2002}
\end{figure}

Generally speaking, there may be a number of broadening mechanisms:

\begin{itemize}

\item Microturbulence, which is formed by small scale (i.e
$\tau \ll 1 $) motions in the atmosphere. In the case of a
supergiant, \Vt\ usualy does not exceed 10 km s$^{-1}$.
In our analysis we determined \Vt\ from a comparison of
observed and computed spectra.

\item Stellar rotation. Our analysis
shows that, in the case of V838, we should adopt \Vr\ = 80 km
s$^{-1}$ to fit the observed spectra. This value is too high for
the later stages of stellar evolution, for obvious reasons.
In reality rotation cannot contribute much to the broadening of
lines observed in spectra of most supergiants.

\item Expansion of the pseudophotosphere of the star.
Asymmetrical profiles of expansion
broadening can be described, to a first approximation, by the formula

$$
G(v,\lambda,\Delta\lambda)= \mbox{const} *
\Delta\lambda\sqrt{1-(\Delta\lambda/\lambda*V_{\rm exp}/c)^2}, \eqno (1)
$$
where \Vexp is the expansion velocity. The observed spectra can be fitted
with \Vexp = 160 km s$^{-1}$ (see Fig. \ref{__profiles}).

\begin{figure}
\begin{center}
\includegraphics [width=88mm, angle=00]{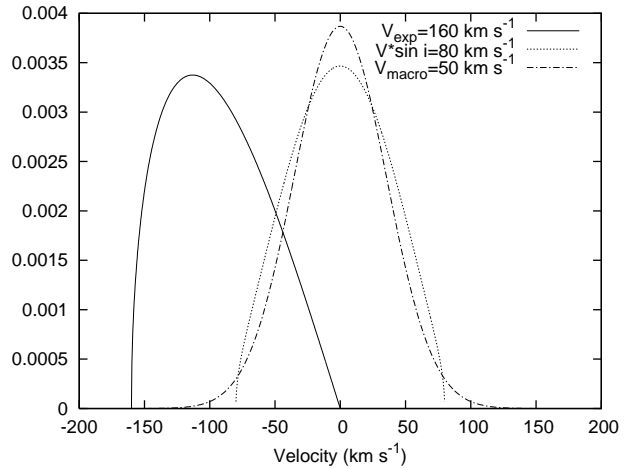}
\end{center}
\caption[]{\label{__profiles} Comparison of expansion (\Vexp =
160 km s$^{-1}$), rotational (\Vr= 80 km s$^{-1}$) and macroturbulent (\Vm= 50
km s$^{-1}$) profiles used in this paper to convolve synthetic spectra.}
\end{figure}

From February 25 to March 2 the brightness of the star was
approximately constant, but at the time of the observation on
March 26 there was a small drop in visual brightness (see Kolev
et al. 2002).
In  Fig. \ref{__sao} we show a comparison of the
observed spectra of V838 Mon at the times of the increase in
luminosity (February 3; Kipper \& Klochkova, private
communication) and at approximately constant luminosity (February
25, March 2 and March 26). They differ drastically -- the spectrum
of February 3 is severely veiled by emission: many lines are
observed in emission.
This demonstrates that effects of the radial expansion
of the line-forming layers were not significant for the dates of
our data and formally obtained value \Vexp = 160 km s$^{-1}$ is
not real.

\item Macroturbulence. After the large increase of luminosity in 2002
January-February,
large scale (i.e. of magnitude $\tau>1$) macroturbulent motions should
be very common in the disturbed atmosphere of V838 Mon. Our numerical
experiments showed that, to get appropriate fits
to the observed spectra taking into account only macroturbulent
velocities, we should adopt \Vmacro $\sim$ 50 km s$^{-1}$.

\end{itemize}

\begin{figure}
\begin{center}
\includegraphics [width=88mm, angle=00]{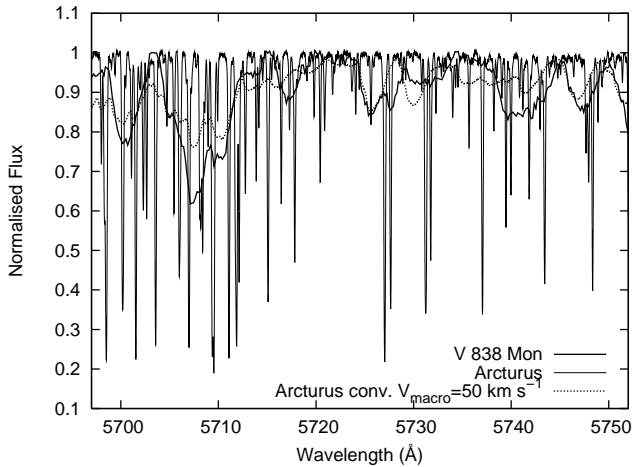}
\end{center}
\caption[]{\label{__Arctur} Comparison of the spectrum of V838 Mon
and that of  Arcturus, convolved with macroturbulent profile \Vm =
50 km s$^{-1}$}
\end{figure}

In any case, for the times of our observations the spectra of V838
Mon resemble the  spectra of ``conventional'' supergiants. Our
V838 Mon spectra for February 25, March 2 and 26 agree, at
least qualitatively, with the spectrum of Arcturus (K2 III),
convolved with macroturbulent velocity profile, given by a
gaussian of half-width \Vmacro = 50 km s$^{-1}$ (Fig.
\ref{__Arctur}).

The observed emissions in the cores of the strongest lines are formed
far outside, perhaps at the outer boundary of the expanding envelope,
i.e. in the region which is heated by shock wave dissipation.

As result of our first numerical experiments, we concluded that
the spectra of V838 Mon in 2002 February - March were similar to
the spectrum of a normal late (super)giant, broadened by strong
macroturbulence motions and/or expansion of its pseudophotosphere.
Unfortunately  we cannot, from the observed spectra, distinguish between
broadening due to the macroturbulence and
expansion (see  next section).

It is worth noting that the
observed spectra of V838 Mon are formed in a medium with
decreasing temperature to the outside, i.e. in the local co-moving system
of co-ordinates the atmosphere, to a first approximation, can be
described by a ``normal'' model, at least in the region of
formation of weak or intermediate strength atomic lines.

\subsection {Fits to observed spectra}

We computed a sample of LTE synthetic spectra for a grid of Kurucz
(1993) model atmospheres with \Tef = 4000 -- 6000 K using the
WITA612 program (Pavlenko 1997). Synthetic spectra were computed
with wavelength step 0.02 \AA, microturbulent velocities 2 -- 18
km s$^{-1}$ with a step 1 km s$^{-1}$, iron abundances $\log
N(\mbox{Fe}) = -5.6\rightarrow-3.6$dex\footnote{in the paper we
use the abundance scale $\sum N_i$ = 1}, with a step 0.1 dex.
Then, due to the high luminosity of the star, we formally adopt
log g = 0. Synthetic spectra were computed using the VALD (Kupka
et al. 1999) line list. For atomic lines the line broadening
constants were taken from VALD or computed following Unsold
(1955).

\begin{figure}
\begin{center}
\includegraphics [width=88mm, angle=00]{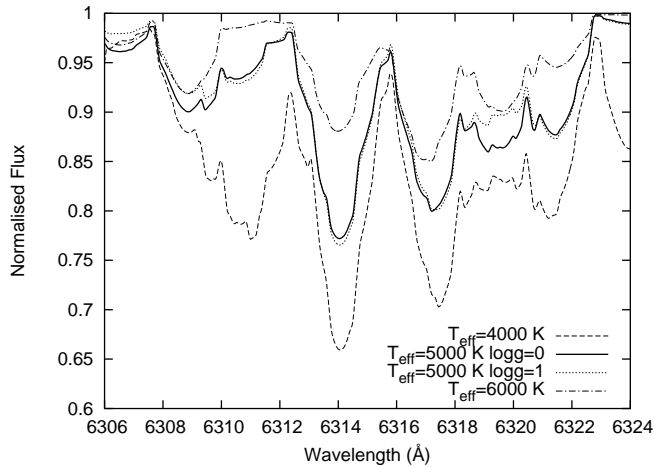}
\end{center}
\caption[]{\label{__Synt__} Dependence of computed spectra on \Tef and log g}
\end{figure}

For the dates of our observations lines of neutral iron dominate
in the spectra. Fortunately, they show rather weak
gravity/pressure dependence, therefore the uncertainty in the
choice of $\log g$ will not be important in determining our
main results; the dependence of the computed spectra on \Tef is
more significant (see Fig. \ref{__Synt__}). The computed
synthetic spectra were convolved with different profiles, and then
fitted to the observed spectra following the numerical scheme
described in Jones et al. (2002) and Pavlenko \& Jones
(2002).

In order to determine the best fit parameters, we compared
the observed residual fluxes $r^{\rm obs}_{\rm \lambda}$ with computed
values $H^{\rm theor}_{\rm \lambda + f_{\rm s}}$. We
let $H^{\rm obs}_{\rm \lambda}$ = $\int F^{\rm
theor}_{x-y}*G(y)*dy$, where  $F^{\rm theor}_{\rm \lambda}$ is
the theoretical flux  and $G(y)$ is the
broadening profile. In our
case $G(y)$ may be wavelength dependent. To get the best fit we
find the minima per point of the 2D function

$S(f_{\rm s},f_{\rm g})=\Sigma(1-H^{\rm synt}/H^{\rm obs})^2$.

We calculated these minimization parameters for our grid of
synthetic spectra to determine a set of parameters $f_{\rm s}$
(wavelength shift parameter) and $f_{\rm g}$ (convolution
parameter).

The theoretical spectra were convolved with a gaussian profile. Our
convolution profile is formed by both expansion and
macroturbulent motions. We cannot distinguish between them in our
spectra. To get a numerical estimate of
the broadening processes in the pseudophotosphere, we use a
formal parameter \Vg, which describes the cumulative effect of
broadening/expansion motions.

The parameters $f_{\rm s}$ and $f_{\rm g}$ were
determined by the minimization procedure; the procedure was
carried out for different spectral regions.
We selected for analysis  6 spectral orders in the interval
5600-6700\AA. In the red, spectral lines are blended by telluric
spectra, and are of lower S/N. In the blue the blending of the
spectra are rather high. Our main intention was to obtain a
self-consistent solution separately for different echelle orders,
and then compare them. If we could obtain similar parameters from
different spectral regions it can be evidence of the reality of
the obtained solution.

\section{Results}

\subsection{The Sun}

To be confident in our procedure, we carried out a similar
analysis for the Sun. For this case we know the solar abundances and other basic
parameters, therefore our analysis provides an
independent estimation of the quality of our procedure:

\begin{itemize}

\item From the solar atlas of Kurucz et al (1984) we extract spectral
regions corresponding to our observed orders of V838 Mon;

\item we convolve the solar spectra with a gaussian of \Vmacro = 50 km s$^{-1}$.

\item we carried out a spectral analysis of the spectral regions
with our procedure; again, model atmospheres from Kurucz (1993)
with a grid of different $\log g$, \Tef, $\log N(\mbox{Fe})$ were
used.

\end{itemize}

\begin{figure}
\begin{center}
\includegraphics [width=88mm, angle=00]{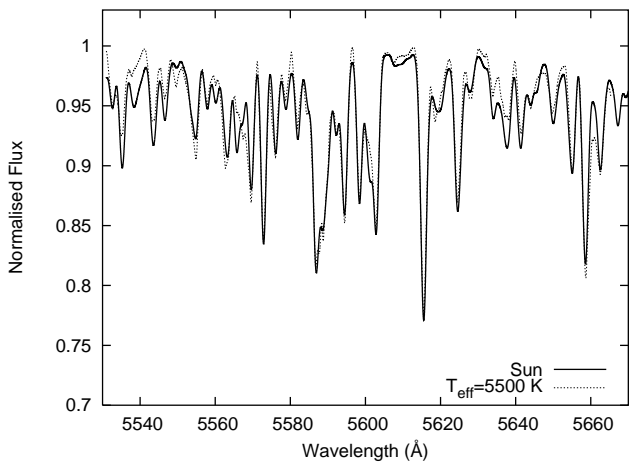}
\end{center}
\caption[]{\label{__sun__} The best fit to the solar spectra.}
\end{figure}

The results of our ``re-determination'' of parameters of the solar
atmosphere are given in Table \ref{__t1__}. The best fit to one
spectral region is shown in Fig. \ref{__sun__}.  From our analysis
of the solar spectrum we obtained \Tef = 5625 $\pm$ 125 K, $\log
N(\mbox{Fe}) = -4.48 \pm 0.15$ dex, \Vt = 1.2 $\pm$ 0.4 km
s$^{-1}$. Here and below we used the standard deviation for
error estimates. All these parameters are in good agreement with
the known parameters of the Sun (Allen 1973) as well as \Vg=
\Vmacro = 44.3 $\pm$ 1.3 km s$^{-1}$ with initial parameter of
convolution.

\begin{table}
\centering
\begin{minipage}{88mm}
\caption{\label{__t1__} Parameters of the solar atmosphere}
\begin{tabular}{@{}llrccclrlr@{}}
 \hline
          Order & Wavelength&       &    \Vt      &         &  \Vmacro \\
            &  range (\AA) &  \Tef (K) & (km s$^{-1}$)& log N(Fe) & (km s$^{-1}$)  \\
 \hline
           &                       &                &   & \\
        11 &      6480 -- 6685       &  5500 &  1    &   -4.5  & 45.8 \\
        12  &     6300 -- 6490       &  5750 &  2    &   -4.6  & 46.4 \\
        13  &     6125 -- 6315       &  5750 &  1    &   -4.4  & 42.9 \\
        14  &     5960 -- 6145       &  5750 &  1    &   -4.2  & 44.1 \\
        16  &     5660 -- 5810       &  5500 &  1    &   -4.6  & 43.9 \\
        17  &     5520 -- 5670       &  5500 &  1    &   -4.6  & 42.9 \\
        &                       &        &     &         \\
     Averaged&                      &  5625 &  1.2  &  -4.48  & 44.3  \\
                &                       &                  &  & \\
 \hline
\end{tabular}
\end{minipage}
\end{table}

\subsection{V838 Mon}

\begin{figure}
\begin{center}
\includegraphics [width=88mm, angle=00]{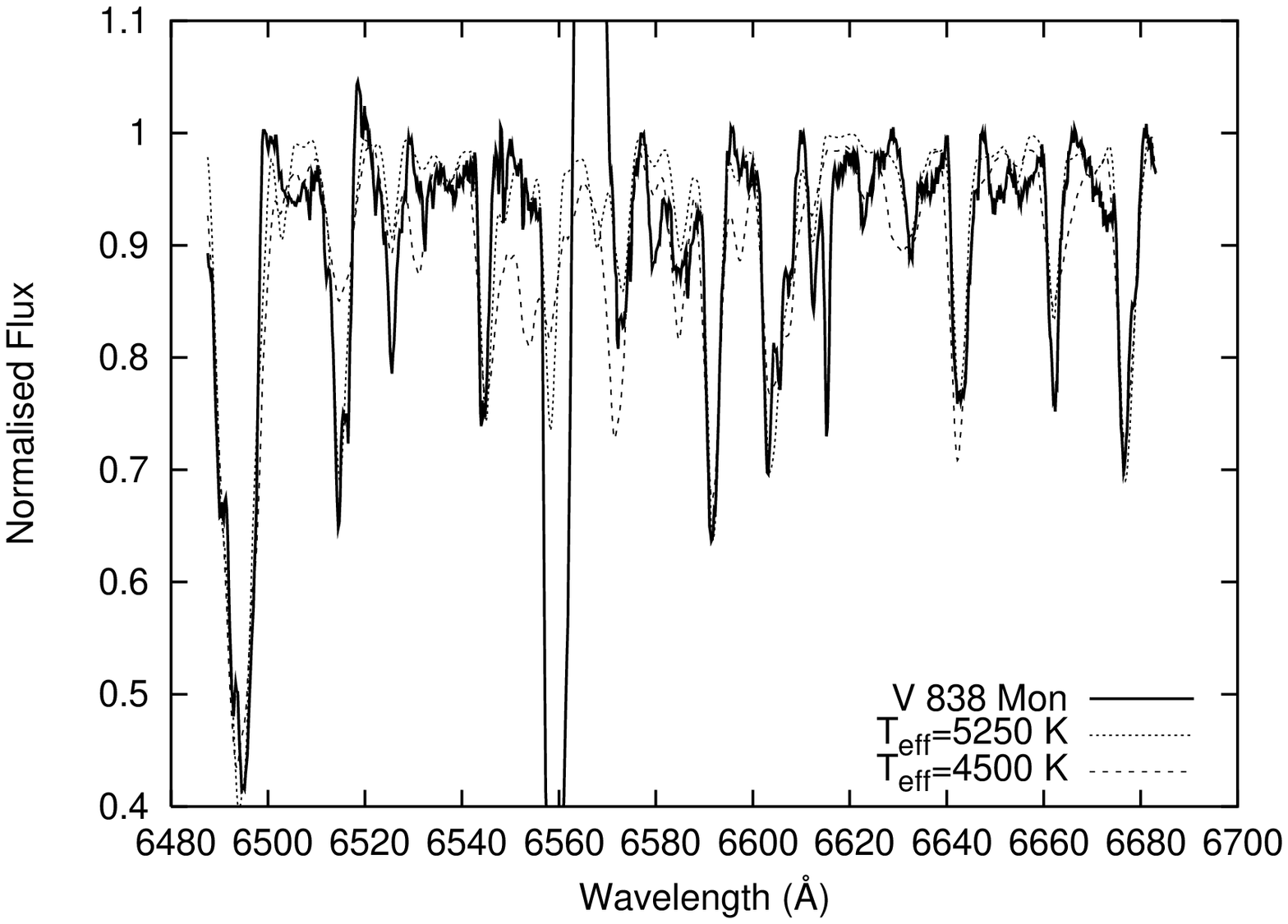}
\includegraphics [width=88mm, angle=00]{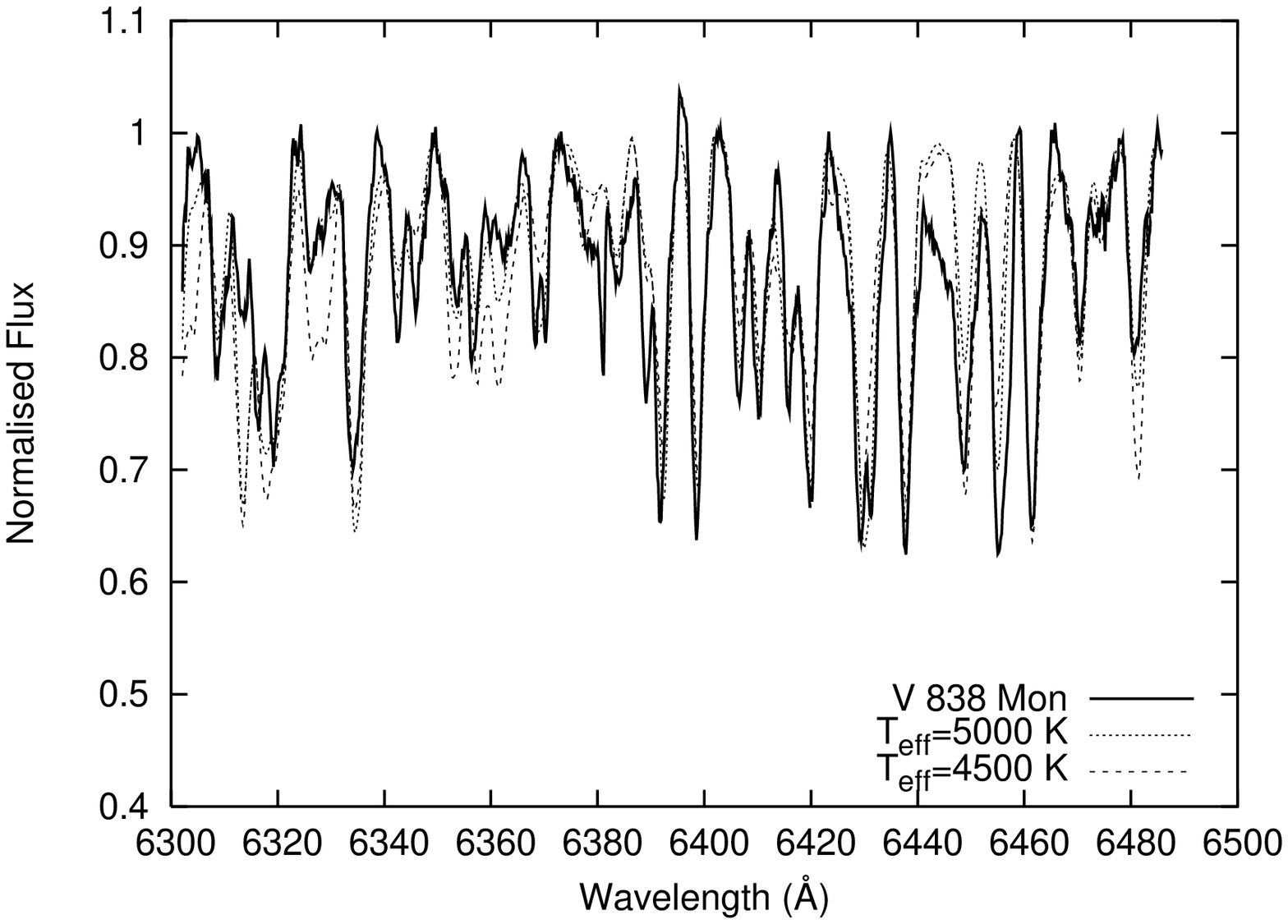}
\includegraphics [width=88mm, angle=00]{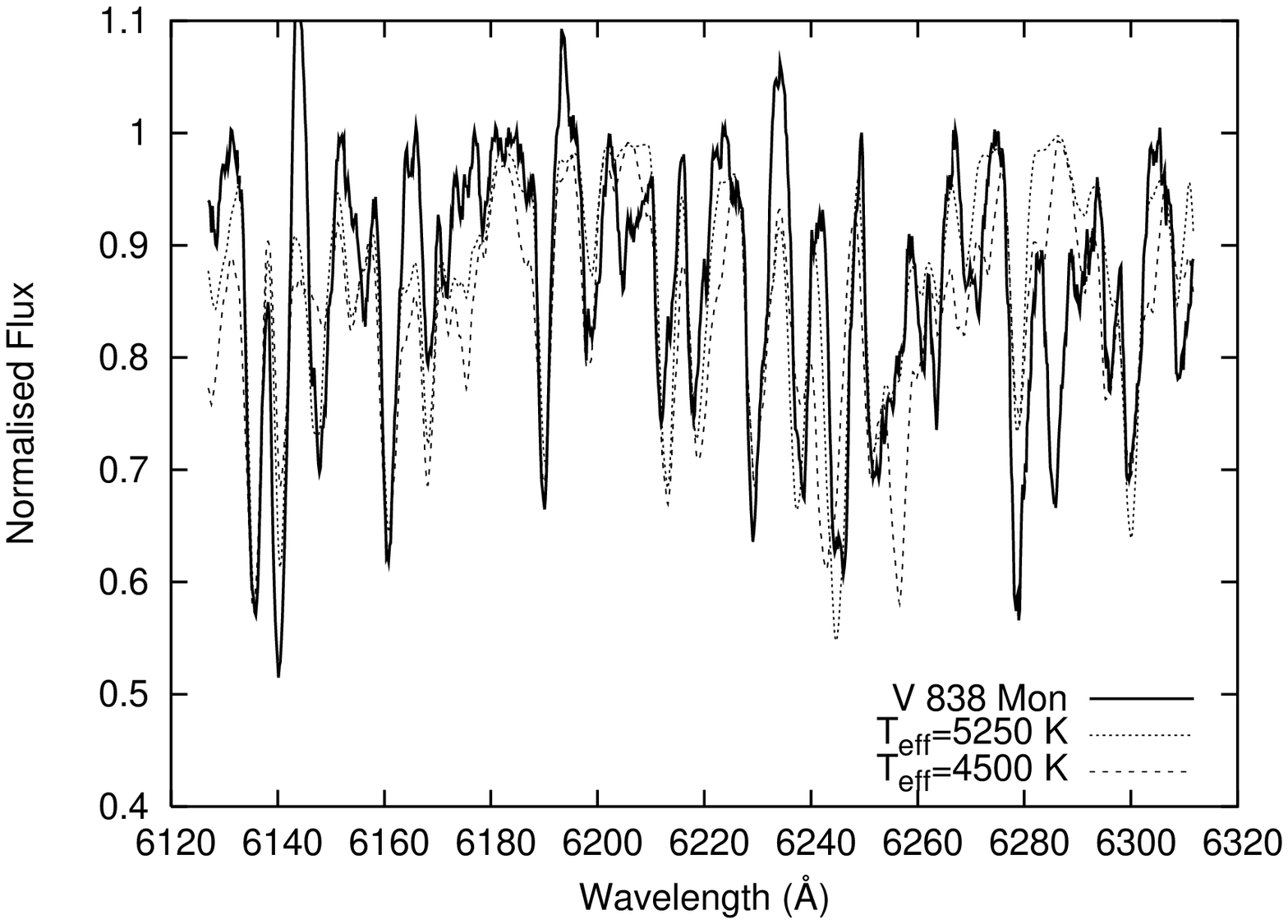}
\end{center}
\caption[]{\label{__best__} The best fits of synthetic spectra to 11-13
orders of the observed spectrum of
 V838 Mon on February 25, found by the minimization procedure.}
\end{figure}

\begin{figure}
\begin{center}
\includegraphics [width=88mm, angle=00]{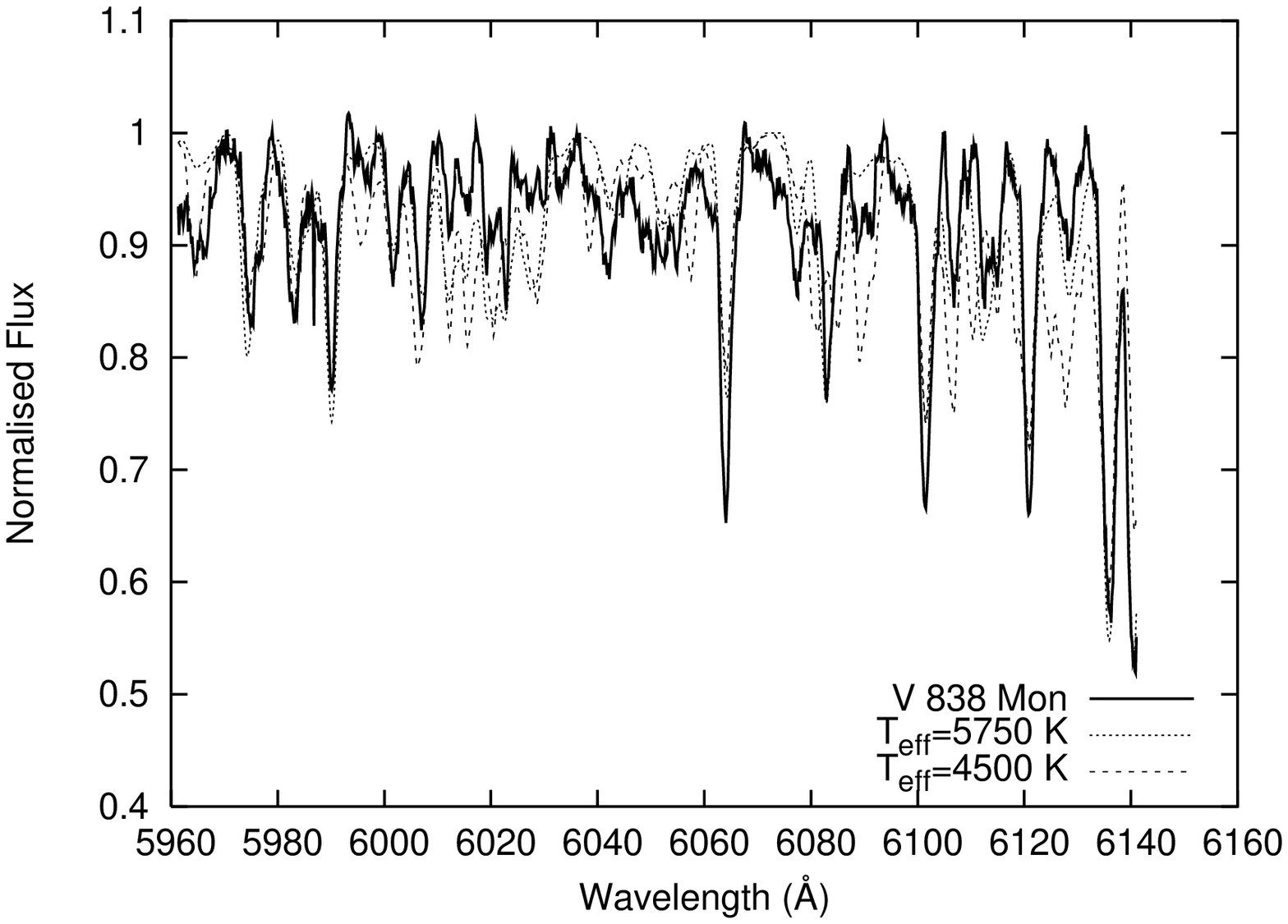}
\includegraphics [width=88mm, angle=00]{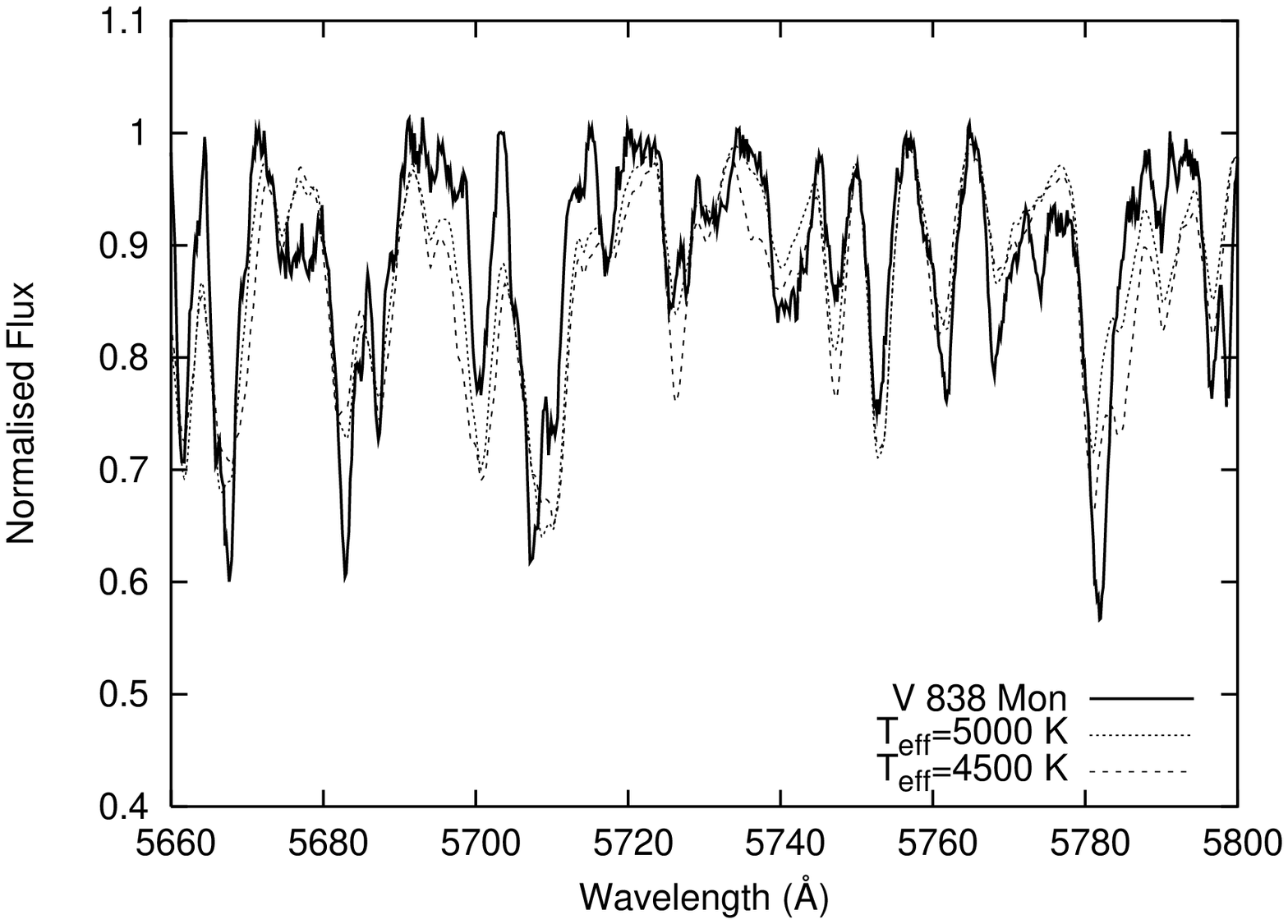}
\includegraphics [width=88mm, angle=00]{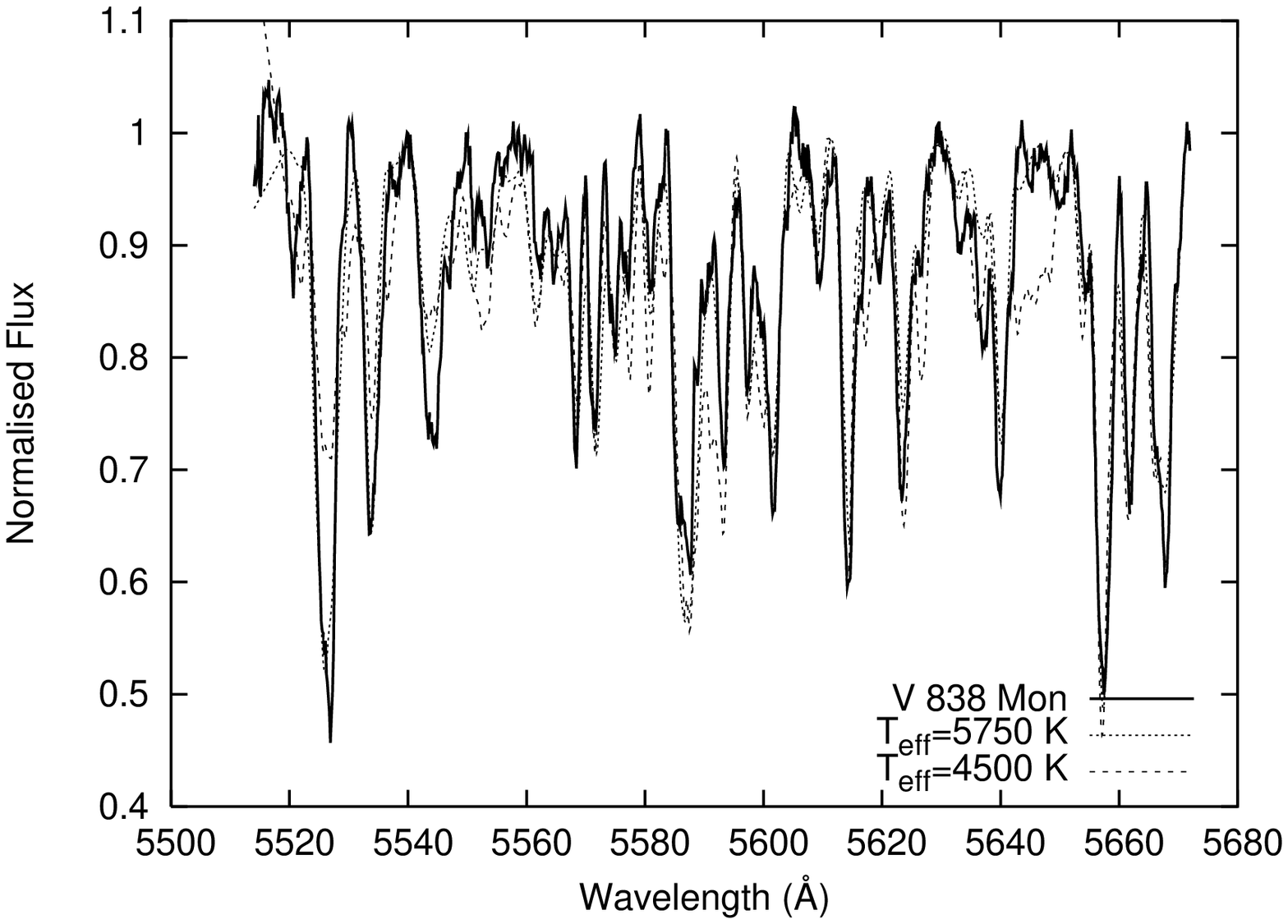}
\end{center}
\caption[]{\label{__best1__} The best fits of synthetic spectra
to orders 14, 16 and 17 of the observed spectrum of V838 Mon on
February 25, found by the minimization procedure.}
\end{figure}

The obtained best fits of the synthetic spectra to the observed
spectrum of V838 Mon for February 25 are shown in Figs.
\ref{__best__} and \ref{__best1__}. Spectral regions containing
emission lines were removed before we conducted  our
analysis. Each plot shows the fits for a) ``the best'' \Tef, and
b)  a case in which the effective temperature has been reduced by
1000 K. The latter value corresponds more with the \Tef
obtained from photometric measurements (c.f. Munari et al. 2002a).
Our procedure allows us to choose the best fit from the grid of
theoretical spectra computed for model atmospheres with different
$\log N(\mbox{Fe})$, \Vt, \Tef. As we see in  Figs.
\ref{__best__} and \ref{__best1__} some differences between
observed spectra and synthetic best fits still remain. In this
work we taken into account only the iron abundance, abundances of
the other elements suggested to be solar. Furthermore, lines of ions
can not be described properly due to too high gravity of used model
atmospheres.

Our main goal was to determine \Tef and $\log N(\mbox{Fe})$ for
the dates of our observations. Results of the determination of the parameters of the pseudophotosphere of
V838 Mon in 2002 February  and  March  are given in Table
\ref{__t2__}.

-- For February 25 we obtained \Tef = 5330 $\pm 300$ K, $\log
N(\mbox{Fe}) = -4.7 \pm 0.14$ dex and \Vt = 13. $\pm$ 2.8 km
s$^{-1}$.

-- For the March 26 data the mean values are \Tef = 4960 $\pm$ 270
K, $\log N(\mbox{Fe}) = -4.68 \pm 0.11$ dex, \Vt = 12.5 $\pm$ 1.7
km s$^{-1}$.

-- And for  March 2 the mean values are \Tef = 5540 $\pm$ 190 K,
$\log N(\mbox{Fe}) = -4.75 \pm 0.14$ dex, \Vt = 13.3 $\pm$ 3.2 km
s$^{-1}$.

-- We obtained \Vg = 54 $\pm$ 3, 47 $\pm$ 3 and 42 $\pm$ 5 km
s$^{-1}$ for February 25, March 2 and March 26, respectively.

--  The $f_{\rm s}$ parameter provides the heliocentric velocity
of V838 Mon. We obtained V$_{\rm radial}$ = $-$76 $\pm$ 3, $-$70
$\pm$ 3 and $-$65 $\pm$ 3 km s$^{-1}$ for February 25, March 2 and
March 26, respectively. Most probably, we see some reduction in
the expansion velocity of the envelope.

\begin{table*}
\centering
\begin{minipage}{140mm}
\caption{\label{__t2__} Atmospheric parameters for V838 Mon}
\begin{tabular}{@{}llcccclrlr@{}}
 \hline
          Order & Wavelength&       &          &          \\
            &  range (A) &  \Tef (K) & \Vt(km s$^{-1}$)& log N(Fe) & \Vg(km s$^{-1}$)  & \Vrad(km s$^{-1}$)  \\
 \hline
    Asiago spectra  &                       &                & & &   \\
                &                       &                  &  & & \\

     February 25  &                       &                & & &   \\
        11 &      6480 -- 6685       &  5250 &  15    &   -4.7 & 53.2 & -79.6  \\
        12  &     6300 -- 6490       &  5000 &  14    &   -4.9 & 54.5 & -76.3  \\
        13  &     6125 -- 6315       &  5250 &  10    &   -4.7 & 56.0 & -82.7  \\
        14  &     5960 -- 6145       &  5750 &  17    &   -4.5 & 52.5 & -79.5  \\
        16  &     5660 -- 5810       &  5000 &  9     &   -4.9 & 60.7 & -67.1  \\
        17  &     5520 -- 5670       &  5750 &  14    &   -4.7 & 51.1 & -73.6  \\
        &                       &        &     &         \\
     Averaged&                     &   5330 & 13.2   &  -4.73 & 54.7 & -76.5  \\
                &                       &                  &  & & \\
     March 26  &                       &                &   \\
            11  & 6480 -- 6685       &  4750&    12   &   -4.8 & 43.7 & -67.3  \\
        12  &     6300 -- 6490       &  4750&    14   &   -4.8 & 44.7 & -68.3  \\
        13  &     6125 -- 6315       &  4750&    10   &   -4.5 & 46.0 & -74.3  \\
        14  &     5960 -- 6145       &  5000&    15   &   -4.8 & 42.1 & -65.8  \\
        16  &     5660 -- 5810       &  5000&    11   &   -4.7 & 38.8 & -52.2  \\
        17  &     5520 -- 5670       &  5500&    13   &   -4.5 & 39.7 & -63.6  \\
        &                       &        &     &         \\
     Average&                       &  4960 &  12.5  &  -4.68 & 42.5 & -65.2  \\
                     &                       &                  & & &  \\
    SAAO spectra  &                       &                & & &   \\

     March 2   &                       &                &   \\
        11  &     6480 -- 6685       &  5500&    12   &   -4.6 & 45.3 & -80.6  \\
        12  &     6300 -- 6490       &  5250&    16   &   -4.9 & 55.8 & -77.3  \\
        13  &     6125 -- 6315       &  5250&    7    &   -4.6 & 42.8 & -78.9  \\
        14  &     5960 -- 6145       &  5750&    14   &   -4.8 & 49.0 & -80.6  \\
        16  &     5660 -- 5810       &  5500&    15   &   -4.9 & 51.7 & -68.1  \\
        17  &     5520 -- 5670       &  6000&    16   &   -4.7 & 42.1 & -80.0  \\
        &                       &        &     &         \\
     Average&                       &  5540&  13.3   &  -4.75 & 47.8 & -77.6  \\
                     &                       &                  &  & & \\
 \hline
\end{tabular}
\end{minipage}
\end{table*}

\section{Discussion}
From a  comparison of our results for all three dates we see that:

\begin{itemize}
\item The effective temperature for March 26 is somewhat lower
then for the previous dates. This is an expected result, in view
of the gradual cooling of envelope. However, for March 2 we found
a slightly higher value of temperature than for February 25. A
possible explanation is the heating of the pseudophotosphere as
result of the third outburst.
\item The microturbulent velocities are very similar and
extremely high for all three dates.
\item Our analysis shows a lower value of \Vg for the later dates: the
effects of expansion and macroturbulence were weakened at the
later stages of evolution of the pseudophotosphere of V838 Mon.
\item The iron abundances $\log N(\mbox{Fe})$ = -4.7 $\pm 0.14$
are similar for all dates.

\end{itemize}

Our estimates of effective temperature are in a good agreement
with Kipper et al. (2004), although we used different procedures
of analysis. The iron abundance ([Fe/H] = $-$0.4 ) and
microturbulent velocity (\Vt = 12 km s$^{-1}$) found by Kipper et
al. (2004) for March 18 are in agreement with our results.

Our deduced ``effective temperatures'' as well as those in
Kipper et al (2004) do not correspond with values obtained from
photometry (\Tef $\sim$ 4200 K). We assume that in our
analysis we deal with temperatures in the line forming region,
rather than with the temperatures at photospheric levels which
determine the spectral energy distribution of V838 Mon and the
photometric indices. Indeed, the formally determined
microturbulent velocity \Vt = 13 km s$^{-1}$  exceeds the sound
velocity in the atmosphere (4-5 km s$^{-1}$). This means that the
region of  formation of atomic lines should be heated by
dissipation of supersonic motions: the temperature there should be
higher than that given in a plane-parallel atmosphere of \Tef
$\sim$ 4200 K.

Certainly the effect cannot be explained by sphericity effects:
the  temperature gradients in the extended atmospheres should be
steeper (see Mihalas 1978), therefore temperatures in the line
forming regions should be even lower, in contradiction with our
results.

Strong deviations from LTE are known to occur during the
photospheric stages of the evolution of novae and supernovae. The
main effect there should be caused by deviations from LTE  in the
ionization balance. However, in our case we used lines of the
neutral iron, which dominate by number. We cannot expect a
reduction in the density of Fe I atoms in the comparatively cool
atmosphere of the star. Furthermore, we exclude from our analysis
strong lines with P Cyg profiles. Lines of interest in our study
have normal profiles.

In our work we used model atmospheres computed by Kurucz (1993)
for the comparative low value of \Vt = 2 km s$^{-1}$. Our estimates of
\Vt in the atmosphere of V838 Mon are much higher. In general,
the temperature structures of the model atmospheres of late spectral
classes respond to changes in \Vt. Nonetheless, in our case, the results
depend mainly on temperatures in the line-forming region. We can use
``standard model atmospheres'' only in the first approach computations.
The dependence of our results on \Vt should be of second order
of importance.

To check this we computed the grid of models with \Vt = 2 and 15
km s$^{-1}$ for \Tef = 4750-5750 K with step 250 K and log g = 1.
Unfortunately we were not able compute models with log g = 0 in
case \Vt = 15 km s$^{-1}$. A program SAM12 (Pavlenko 2003) was
used. Line opacity treatment was used to account absorbtion of
atomic and ion lines. Then we carry out our fiting procedure for
this grid of models for February 25 data. We found that
differences in results less then 0.1 dex in iron abudance and 1 km
s$^{-1}$ in \Vt. Still this ``\Vt factor'' must be considered in
detail in future work based on the more sophisticated models.

In the present work we determined only the iron abundance. The
effects of other element abundances should be taken into account
as well. However, the obtained effective temperatures of V838 Mon
are rather low and lines of neutral iron are of prime importance.
Changes in the abundances of other  free electrons donors --
sodium, potassium, etc. -- cannot affect the ionization
equilibrium of iron. However, we suggest continued attempts to
determine their abundances. In addition to their direct importance
to the determination of the evolutionary status of V838 Mon, they
are also of importance for the determination of the opacities in
the stellar atmosphere. Generally, to determine the abundances of
a few elements from modelling of complex blends we need to know a)
the number and b) relative intensities of contributing lines, i.e
the abundances of all elements.

\section{Conclusions}

We have carried out our study in the framework of a comparatively
simple model. The static LTE model for the star with
expanding pseudophotosphere is the rough enough approximation.
Most of our results are rather qualitative, and they should be
confirmed and/or refined in the future. However, in general we are
confident enough in our results, in particular:

\begin{itemize}

\item we obtain some evidence of cooling of the line-forming region;

\item most probably, the line-forming region is heated by
supersonic motions -- our spectroscopic temperatures exceed
photometrically determined \Tef by $\sim$ 1000 K;

\item we do not find any significant change in the iron abundance in
atmosphere V838 from February 25 to March 26.

\item we derived a moderate deficit of iron $\log N(\mbox{Fe})$ $\sim$ $-$4.7
in the atmosphere of V838 Mon.

\end{itemize}

\section*{Acknowledgments}
We thank Drs. Ulisse Munari, Lisa Crause, Tonu Kipper and
Valentina Klochkova for providing  spectra and for discussions of
our results. We thank Dr. Nye Evans for improving text of paper.
We thank unknown referee for many helpful remarks. This work
was partially supported by a PPARC visitors grants from PPARC and
the Royal Society. YPs studies are partially supported by a Small
Research Grant from American Astronomical Society.

This research has made use of the SIMBAD database,
operated at CDS, Strasbourg, France.

\bsp

\label{lastpage}

\end{document}